\documentclass[11pt]{article}

\usepackage{latexsym}
\usepackage{amsmath}
\usepackage{float}
\usepackage{bm}
\usepackage{amssymb}
\usepackage{xcolor}
\usepackage{amsthm}
\usepackage{dsfont}
\usepackage[ansinew]{inputenc}
\usepackage{graphicx}
\usepackage{epsfig}
\usepackage{dsfont}
\usepackage{bbm}
\usepackage{url}
\usepackage{placeins}
\usepackage{multirow}
\usepackage{algorithm} 
\usepackage{enumitem}
\usepackage{overpic}


\usepackage[colorlinks]{hyperref}
\hypersetup{allcolors =blue}

\usepackage[authoryear]{natbib}
\renewcommand{\epsilon}{\varepsilon}

\newcommand{\EE}{\mathds{E}}

\begin{document}

\title{{\bf \Large 
Capturing heterogeneous time-variation in covariate effects in non-proportional hazard regression models
%Capturing treatment- or subgroup-specific time-variation in covariate effects in non-proportional structured hazard regression models
%Capturing treatment-specific time-variation in Cox-type structured hazard regression using heterogeneously varying coefficients
}}
\author{Niklas Hagemann$^{1,2}$, Thomas Kneib$^{3}$ and
        Kathrin M\"ollenhoff$^{1,2}$  \\
 \small{$^{1}$ Institute of Medical Statistics and Computational Biology, Faculty of Medicine, University of Cologne,}  \\ \small Germany  \\
 \small $^{2}$ Division of Mathematics, Department of Mathematics and Computer Science, University of Cologne, Germany  \\
 \small{$^{3}$ Chair of Statistics and Campus Institute Data Science, Georg August University G\"ottingen, Germany
 } }

\maketitle

\begin{abstract}
A central focus in survival analysis is examining how covariates influence survival time. These covariate effects are often found to be either time-varying, heterogeneous -- such as being specific to patients, treatments, or subgroups -- or exhibit both characteristics simultaneously.
    %One of the major topics in survival analysis is analyzing the effect covariates have on the survival time. Frequently, these covariate effects are observed to be either time-varying, heterogeneous, i.e. patient-, treatment- or subgroup-specific, or even both. 
    While the standard model, the Cox proportional hazards model, allows neither time-varying nor heterogeneous effects, several extensions to the Cox model as well as alternative modeling frameworks have been introduced. However, no unified framework for incorporating heterogeneously time-varying effects of covariates
    has been proposed. %none of these studies includes heterogeneously time-varying effects of covariates. 
    Such effects occur when a covariate influences survival not only in a heterogeneous and time-varying manner, but when the time-variation is also heterogeneous. 
    
    We propose to model such effects by introducing heterogeneously time-varying coefficients to piecewise exponential additive mixed models. We deploy functional random effects, also known as factor smooths, to model such coefficients as the interaction effect of heterogeneity and time-variation. Our approach allows for non-linear time-effects due to being based on penalized splines and uses an efficient random effects basis to model the heterogeneity. Using a penalized basis prevents overfitting in case of absence of such effects. In addition, the penalization mostly solves the problem of choosing the number of intervals which is usually present in unregularized piecewise exponential approaches. We demonstrate the superiority of our approach in comparison to competitors by means of a simulation study. Finally, the practical application and relevance are outlined by presenting a brain tumor case study.
\end{abstract}

\section{Introduction} \label{sec:intro}
One of the major topics in survival analysis is analyzing the effect covariates have on the survival time. Frequently, the effects of these covariates can be observed to be either time-varying, heterogeneous, i.e. patient-, treatment- or subgroup-specific, or even both.
If the goal is to analyze the effects of covariates on the survival time, hazard regression models play a critical role. 
%Hazard regression models 
They estimate the hazard function, which represents the instantaneous rate of occurrence of the event at a given time, conditional on survival up to that time. By incorporating covariates, hazard regression models allow to assess the impact of various factors on the hazard rate and, hence, to identify significant effects.
The widely used standard model is the \emph{Cox proportional hazards model} \citep{cox1972} where the hazard rate of an observation $i \in \{1,...,n\}$ with corresponding covariate vector $\boldsymbol{x}_i $ %\in \RR^P$ 
is given by
$$
\lambda_i(t)=\lambda_0(t) \exp(\boldsymbol{x}_i^\top \boldsymbol{\beta}),
$$
where $\boldsymbol{\beta}$ is the vector of regression coefficients. 
The assumption of proportionality of the hazards results from the model being strictly split into the time-dependent baseline hazard $\lambda_0(t)$ and the time-constant covariate effects $\exp(x_i^\top \beta)$.
In addition, the Cox model assumes the covariate effects to be (exp-transformed) linear effects.

% The fundamental assumption of proportionality of the hazards results from the model being strictly split into two multiplicative parts: the time-dependent baseline hazard $\lambda_0(t)$ and the time-constant covariate effects $\exp(x_i^\top \beta)$. This does not only imply that the coefficients are time-constant but also that the covariates do not vary over time.
% In addition, the Cox model assumes the covariate effects to be (exp-transformed) linear effects. 

%However, several review paper \citep{Kuitunen2021, Ngari2021, Altman1995} show that in applied research this assumption is often not checked for compliance.
%As outlined before, t
These strict assumptions are often not fulfilled in 
practice \citep{Li2015, Jachno2019}. 
%and several review paper \citep{Kuitunen2021, Ngari2021, Altman1995} show that the assumptions are often not even checked for compliance.
Frequently, this is caused by non-proportional hazards, i.e. the covariates or their effects are time-dependent. 
In addition, in many cases the effects might not be linear but of a more complex form. 
In particular, heterogeneity in terms of treatment-specific, subgroup-specific (e.g. gender-specific) or individual effects might be present which cannot be captured properly by linear effects. % but needs the introduction of random effects.
%However, these assumptions are quite strict and often not fulfilled in partial research. 
Therefore, several flexible extensions to the Cox model have been introduced. On the one hand, such extensions relax the proportional hazards assumption: 
The most commonly used may be the stratified Cox model in which for each level of a categorical variable a separate baseline hazard is fitted. Alternatively, several studies \citep[see, e.g.,][]{zucker1990, murphy1991} propose to include time-varying coefficients in order to capture time-dependent covariate effects. %, i.e. time-varying coefficients. 
In addition, \citet{andersen1982} introduced an approach to include time-dependent covariates.% and several approaches \citep[see, e.g.][]{zucker1990, murphy1991} introduced time-dependent covariate effects, i.e. time-varying coefficients. 
%Another way to address violations of the proportional hazards assumption is to use the stratified Cox model \citep{} in which for each level of the categorical variable a separate baseline hazard is fitted. 

On the other hand, extensions of the Cox model have been introduced to allow for more complex effects: %ZZZ included non-linear spline-based baseline hazards and XYZ 
\citet{Gray1992} added non-linear smooth spline-based covariate effects and \citet{Hess1994} uses such effects to express covariate effects as a function of time. 
%Regarding the heterogeneity is often present in medical research, eighter between individual participants, subgroups (e.g. between genders) or treatments. 
Regarding the heterogeneity, again the stratified Cox model can be mentioned, where the baseline hazard can be group-specific.
However, a more natural way to account for heterogeneous effects, which also allows for more general types of heterogeneity, is to introduce random effects leading to \emph{frailty models} \citep{vaupel1979, ripatti2000, therneau2003}.
%An overview of such extensions to the cox model can be found in 

In contrast to adding specific extensions to the Cox model, several recent studies have aimed to introduce a new flexible hazard regression framework: \citet{kneib2007} introduce \emph{Cox-type structured hazard regression models}
\begin{equation} \label{eq:shrm}
        \lambda_i(t|\boldsymbol{x}_i) =\lambda_0(t) \exp \left( \sum_{k=1}^K f_k(\boldsymbol{x}_i, t)\right) =\exp \left( \Tilde{\lambda}_0(t) + \sum_{k=1}^K f_k(\boldsymbol{x}_i, t)\right),
\end{equation}
where $\Tilde{\lambda}_0(t)$ is the log-baseline hazard and $f_k$ can resemble different types of effects, e.g. linear effects, smooth (spline-based) effects, time-varying effects or random effects/frailty. The inclusion of time-varying effects allows for explicit modeling of non-proportional hazards. This approach was further investigated by \citet{hofner2011, hofner2013}. The corresponding inference is conducted by mixed model-based penalized likelihood estimation. However, the log-likelihood involves an integral over the hazard rate. Hence, the estimation relies on numerical integration which is computationally costly and can be subject to impreciseness. Alternatively, \citet{hennerfeind2006} proposed a Bayesian estimation
scheme for such models. However, this approach is based on the same log-likelihood and, therefore, shares these disadvantages. 

Another approach to introduce models of the form \eqref{eq:shrm} is given by \emph{piecewise exponential additive mixed models} \citep[PAMM;][]{bender2018a, bender2018b} which generalizes the concept of \emph{piecewise exponential models} \citep[PEM;][]{friedman1982} from linear to additive predictor terms.
The underlying idea is to divide the time axis into a finite
number of intervals and assume the hazard rate to be piecewise constant within these intervals. 
While manually choosing the interval cut-off points is a challenging task and a frequent source of criticism for PEMs, PAMMs avoid the arbitrary choice of cut-off points by using a penalized approach providing a sufficiently good fit while preventing overfitting.
Under the assumption of piecewise constant hazard rates, restructuring the data leads to the likelihood of the survival model being proportional to the one of a Poisson regression model (see Section 2 of \citet{bender2018a} for details).
Hence, both models are equivalent with respect to their maximum likelihood estimators and the model parameters are estimated based on the Poisson model. 
Therefore, estimation can make use of existing methods and implementations for generalized linear models (GLMs) in the case of PEMs and generalized additive (mixed) models (GAMs/GAMMs) in the case of PAMMs. 
%Several sudies () used this for more specific models but PAMMs give are the most general and comprehensive framework for such models.   

While these approaches %\citet{kneib2007}, \citet{hennerfeind2006}, \citet{bender2018a} 
propose several flexible effects such as non-linear, time-varying and random effects, none of them  
%However, none of these studies 
include heterogeneously time-varying effects of covariates. Such effects occur if a covariate influences the survival time not only in a heterogeneous and time-varying manner, but the time-varying effect is heterogeneous, too. 
A typical example would be that the effect of a covariate is treatment specific, time-varying and that its time-variation is also treatment-specific, e.g., decreasing for an intervention but increasing for a placebo. 
To the best of our knowledge, this study is the first to propose heterogeneously time-varying covariate effects in hazard regression models.

%Based on this modeling framework, 
Based on the framework of PAMMs, 
we introduce these heterogeneously time-varying coefficients as 
$$
f_{kg}(t) \cdot x_{ik}, %\, k\in \{1,...,K\}
$$
where $g \in \{ 1,...,G\}$ is the grouping variable. Besides treatments, the grouping can also correspond to characteristics of the participants (e.g. gender or subdiagnoses), characteristics of the study (e.g. centers in multicenter studies) or even individual heterogeneity. 
We propose to model such heterogeneously time-varying coefficients based on 
%\citet{hagemann2024} propose to capture heterogeneity in (time-) varying coefficients by using 
\emph{functional random effects} \citep[FRE;][]{kneib2019}. %as varying coefficients. 
This leads to \emph{functional random coefficients}, which \citet{hagemann2024} recently proposed to use to capture heterogeneous time-variation in covariate effects.
%within an other model class, namely conditional logit models which are popular in market research.  
Their study focuses on conditional logit models, a class of models commonly used in marketing research, but the approach is directly generalizable to other GAMs and can therefore be applied to PAMMs as well. 
%Such heterogeneously (time-) varying coefficients can be introduced by using \emph{functional random effects} \citep[FRE;][]{kneib2019} as varying coefficients. 
Functional random effects are also known as \emph{factor smooth interactions} or \emph{random wiggly curves} \citep[][]{wood2017} and are essentially tensor product interactions of smooth effects and random effects. %Therefore, they induce anisotropy and are quite flexible with regard to the underling main effects, the choice of basis functions and penalty terms.
%To use FREs in order to capture heterogeneity in (time-) varying coefficients has been recently proposed by \citet{hagemann2024} in conditional logit models which are popular in market research.  

This paper is structured as follows: 
In Section \ref{sec:model}, piecewise exponential hazard regression models including the corresponding inference are succinctly discussed. 
In Section \ref{sec:FRE}, we introduce subgroup-specific time variation in covariate effects using functional random coefficients. 
%In Section \ref{sec:FRE}, we introduce functional random coefficients in order to capture subgroup-specific time-variation in covariate effects. 
A simulation study is conducted in Section \ref{sec:simu} to show the ability of our approach to capture these effects as proposed. In addition, the simulations demonstrate that the penalized approach prevents overfitting in the absence of such effects.
Section \ref{sec:case} illustrates the method and outlines its practical relevance by investigating the effect of fraction genome altered as a predictor of survival time in patients with brain tumors. 
%using a brain tumor survival example before 
Finally, Section \ref{sec:conclusion} closes with a discussion.

%\section{Heterogeneously time-varying effects in structured additive hazard regression models} 
\section{Piecewise exponential hazard regression models}\label{sec:model}
\subsection{Piecewise exponential models}\label{sec:model:PEM}
%We base our modeling approach on the framework of piecewise exponential additive mixed models. That is, assuming 

Piecewise exponential models \citep[PEM;][]{friedman1982} are an alternative to classical approaches in survival regression, especially to the Cox model. Their main advantage is that the corresponding inference can be based on a Poisson model and, hence, can make use of existing tools for generalized linear models (GLMs) and generalized additive models (GAMs). They require the partition of the time axis into a finite number of intervals and assume the hazard rate to be constant within each interval.
% In order to introduce PEMs formally, let $T_i$ denote the true survival time and $C_i$ the (non-informative) censoring time for subject $i$ such that $t_i := \min(T_i, C_i)$ is its observed right-censored time under risk. 
The piecewise exponential model is %then 
defined as 
\begin{equation} \label{eq:pem}
    \lambda_i(t|\boldsymbol{x}_i) = \lambda_0(t_j) \exp(\eta(\boldsymbol{x}_i, t_j)) \quad \forall \, t \in (\kappa_{j-1}, \kappa_{j}],
\end{equation}

where $\eta(\boldsymbol{x}_i, t_j)$ is the predictor term, 
$(\kappa_{j-1}, \kappa_{j}], \, j =1,...,J$ are the intervals for which the hazard rate is assumed to be constant, $\kappa_0 = 0$ and $\kappa_J = \max(t)$. There are different ways of choosing $t_j, j=1,...,J$, i.e. the time values at which the hazard function \ref{eq:pem} is evaluated. % for all 
%within the intervals, where the 
The two most frequently used approaches are interval end-points $t_j = \kappa_j \, \forall \, t \in (\kappa_{j-1}, \kappa_{j}]$ and interval mid-points $t_j = 0.5(\kappa_j + \kappa_{j-1}) \, \forall \, t \in (\kappa_{j-1}, \kappa_{j}]$. 

In order to make use of the piecewise exponential approach, it is convenient to restructure the data as outlined by \citet{friedman1982} and \citet{bender2018a}. 
Therefore, let $T_i$ denote the true survival time and $C_i$ the (non-informative) censoring time for subject $i \in \{1,...,n\}$ such that $t_i := \min(T_i, C_i)$ is its observed right-censored time under risk. 
The data is then restructured, such that for each subject $i$ there is a row for each interval $j$ in which it was under risk. 
These rows contain $t_{ij}$, 
%as well as a 
an interval-specific event indicator $\delta_{ij}$, formally being defined by
$$
\delta_{ij} = \left\{ \begin{array}{ll}
     1 \quad &  \text{if }t_{i} \in (\kappa_{j-1}, \kappa_{j}] \text{ and } t_i = T_i,\\
     0 \quad & \text{else, }
\end{array} \right.
$$
%In addition, each row contains 
as well as an offset value $o_{ij}= \log(t_{ij})$ that gives the log-transformed time under risk and will be needed for the model estimation. 
%This data transformation can be conducted in \texttt{R} by using the function \texttt{as\_ped} from the package \texttt{pammtools} \citep{bender2018b}. 

\citet{friedman1982} proposed a linear time-constant predictor $\eta(\boldsymbol{x}_i, t_j)= \boldsymbol{x}_i^\top \boldsymbol{\beta}$ implying a proportional hazards model. However, this easily generalizes to more complex effects, including time-varying effects, which implies a non-proportional hazards model. This leads to the class of piecewise exponential additive mixed models \citep[PAMM;][]{bender2018a, bender2018b}. 

% The underlying assumption of piecewise exponentially distributed survival times 
% %models are based on the assumption that the survival times 
% %This 
% implies a piecewise constant hazard rate 
% %is piecewise constant, i.e.
% $$
% \lambda_i(t) = \lambda_i(t_j) \quad \forall \, t \in (\kappa_{j-1}, \kappa_{j}]
% $$
% with $J$ intervals $(\kappa_{j-1}, \kappa_{j}], \, j =1,...,J$ for which the hazard rate is assumed to be constant, $\kappa_0 = 0$ and $\kappa_J \geq \max(t)$. There are different ways of choosing $t_j$ within the interval and the two most frequently used approaches are interval end-points $t_j = \kappa_j \, \forall \, t \in (\kappa_{j-1}, \kappa_{j}]$ or interval mid-points $t_j = 0.5(\kappa_j + \kappa_{j-1}) \, \forall \, t \in (\kappa_{j-1}, \kappa_{j}]$. In addition, we assume non-informative right censoring. 

% In order to make use of the piecewise exponential model, the data needs to be restructured as outlines by \citet{friedman1982} and \citet{bender2018a} such that for a subject $i$ there is a row for each interval in which it was under risk. These rows contain $t_j$, 
% %as well as a 
% an interval-specific event indicator $\delta_{ij}$, formally being defined by
% $$
% \delta_{ij} = \left\{ \begin{array}{ll}
%      1 \quad &  \text{if }t_{i} \in (\kappa_{j-1}, \kappa_{j}] \text{ and } t_i = T_i,\\
%      0 \quad & \text{else, }
% \end{array} \right.
% $$
% %In addition, each row contains 
% as well as an offset value $o_{ij}= \log(t_{ij})$ that gives the log-transformed time under risk and will be needed for the model estimation. 

\subsection{Piecewise exponential additive mixed models}\label{sec:model:PAMM}
% We base our modeling approach on the framework of piecewise exponential additive mixed models. That is, assuming that the hazard rate is piecewise constant, i.e.
% $$
% \lambda_i(t) = \lambda_i(t_j) \, \forall \, t \in (\kappa_{j-1}, \kappa_{j}]
% $$
% with $j =1,...,J$ intervals $(\kappa_{j-1}, \kappa_{j}]$ for which the hazard rate is assumed to be constant, $\kappa_0 = 0$ and $\kappa_J \geq \max(t)$. There are different ways of choosing $t_j$ within the interval and the two most frequently used approaches are interval end-points $t_j = \kappa_j \, \forall \, t \in (\kappa_{j-1}, \kappa_{j}]$ or interval mid-points $t_j = 0.5(\kappa_j + \kappa_{j-1}) \, \forall \, t \in (\kappa_{j-1}, \kappa_{j}]$

%Under this assumption, 
Using PAMMs, a structured additive hazard regression model of the form \eqref{eq:shrm} can be constructed as
$$
    \lambda_i(t|\boldsymbol{x}_i) =
    %\lambda_0(t_j) \exp \left( \sum_{k=1}^K f_k(\boldsymbol{x}_i, t_j)\right) =
    \exp \left( \Tilde{\lambda}_0(t_j) + \sum_{k=1}^K f_k(\boldsymbol{x}_i, t_j)\right), \quad \forall \, t \in (\kappa_{j-1}, \kappa_{j}].  
$$
where $\boldsymbol{x}_i$ denotes the covariate vector for subject $i$. % \in \RR^P$ denotes the covarite vector for person $i$. As usual for GAMs, the effect index $k=1,...,K$ is not the same as the covariate index $p=1,...,P$ since one effect can consist of several variables (e.g. interaction effects) and one variable can be contained in several effects (e.g. as main and interaction effect). 
The above notation slightly deviates from the one of \cite{bender2018a} as we do not include different effect types explicitly
%like, for example, time constant effects or frailty but
but implicitly as special cases of $f_k(\boldsymbol{x}_i, t_j)$. Typical examples are, among others, linear effects $f_k(\boldsymbol{x}_i, t_j) = \beta_p \cdot x_{ip}$, time-constant non-linear effects $f_k(\boldsymbol{x}_i, t_j) = f_k(\boldsymbol{x}_i)$, linearly time-varying effects $f_k(\boldsymbol{x}_i, t_j) = \beta_p \cdot  x_{ip} \cdot  t_j$ as well as frailty/random effects. 
In addition, this can be easily generalized to time-varying covariates $x_{ipt}$. 
% since the models are estimated as GAMs. 
However, since time-varying covariates are not our focus, we omit them here for notational simplicity and refer the reader to section 3.4 of \citet{bender2018a}. 
%this also includes time-varying covariates, e.g. $f_k(\boldsymbol{x}_i, t_j) = f_k(\boldsymbol{x}_i(t))$.

In contrast to classical PEMs, for which \citet{friedman1982} suggested a step
function as baseline hazard, \citet{bender2018a} propose to use a penalized regression spline as baseline hazard for PAMMs.
This eliminates the problem of manually selecting the interval cutoff points, which is a challenging task and a common criticism of PEMs.
%, PAMMs avoid the arbitrary choice of cut-off points by using a penalized approach. 
By deploying a penalized approach, the number of cut-off points just needs to be large enough to provide a sufficiently good fit while overfitting is prevented due to the penalization. Hence, the standard choice of cut-off points for PAMMs is using all unique observed survival times. 
For sufficiently large datasets with relatively dense and %survival times and 
precisely measured survival times (leading to only few ties) this usually leads to very narrow intervals. 
%Offnote, in case of time-varying covariables the change points of these covariates need to be included in the cut-off points aswell %at which (changes in) the time-dependent covariate(s) were recorded \citep{bender2018b}.
Therefore, the assumption of constant hazard rates is not very strict in practice, as it only applies for quite short intervals. 
In addition, in many applications this makes the choice of $t_j$ within the interval mostly irrelevant as there are no large differences. Hence, interval end points, i.e. $t_j = \kappa_j$, are often just chosen for simplicity. 

%providing a sufficiently good fit while preventing overfitting.
% It is worth emphasizing that time-constant effects are implicitly included as special case via $f_k(\boldsymbol{x}_i, t_j) = f_k(\boldsymbol{x}_i)$ also include time-constant effects as special cases

\subsection{Poisson-likelihood  based inference and software implementation} \label{sec:model:inference}
As outlined by \citet{bender2018a} the main advantage of PAMMs is that their likelihood is proportional to the one of a Poisson regression model
\begin{equation} \label{eq:poisson_model}
    \EE(\delta_{ij}|x_{i}) = \exp(\Tilde{\lambda}_0(t_j) + \eta(\boldsymbol{x}_i, t_j) + o_{ij}).
\end{equation}
% which is given by 
% $$
% \log(L(\boldsymbol{\beta}| \boldsymbol{y}, \boldsymbol{X}, t)) = \sum_{i=1}^n \left( y_i \eta_i - \exp(\eta_i) - \log(y_i!) \right),
% $$
Hence, both models are equivalent with respect to their parameters and the corresponding estimation can be conducted based on the Poisson model. 
%Due the Poisson model being a GAM, 
Therefore, the inference can be based on existing methods %for such models. Therefore, 
and one can make use of the methodological and algorithmic advances in the estimation of GAMs. 
This includes both, frequentist \citep[e.g.][]{wood2011} as well as Bayesian methods \citep[e.g.][]{kneib2019}. 
While the Bayesian approach may have theoretical and interpretive advantages in many cases, it is often not numerically feasible because %due to the enlargement of the datasets due to 
the data transformation discussed in Section \ref{sec:model:PEM} can strongly enlarge the datasets leading to very high computational costs.

Estimating the smoothing parameters is a challenging task in frequentist inference. 
Besides other alternatives (see, e.g. \citet{Fahrmeir2022} or \citet{wood2017} for an overview), \citet{wood2011} proposes a method utilizing a random effects perspective while avoiding the formal mixed model framework. 
The smoothing parameters are estimated directly from the restricted likelihood function without requiring the specification of a full mixed model structure. 
This is achieved by using a direct Laplace approximation that integrates out the random effects, i.e. the spline coefficients. Hence, this method optimizes a well-defined likelihood function directly with respect to the smoothing parameters. Therefore, it bypasses the need to solve mixed-model equations.

As discussed by \citet{wood2017}, this method is advantageous compared to smoothness selection criterion-based and full mixed model-based approaches in terms of convergence, precision and numerical stability.

\section{Capturing heterogeneous time-variation in covariate effects using functional random coefficients} \label{sec:FRE}
Within the framework of PAMMs several effect types have already been introduced, including smooth, linearly time-varying effects $t_j \cdot f_k(x_{ik})$, linear, smoothly time-varying effects $f_k(t_j) \cdot x_{ik}$, smooth, smoothly time-varying effects
$f_k (x_{ik} , t_j)$ (see table 3 of \citet{bender2018a} and table 1 of \citet{bender2018b} for a complete  overview) as well as random effects (in terms of log-normal frailty). 
However, heterogeneously time-varying effects of covariates
have not yet been considered. They can be denoted as  
$$
f_{kg}(t) \cdot x_{ik},
$$
where $g \in \{ 1,...,G\}$ is the grouping variable. Formally, $g$ also has an index $i$ but we omit that here to avoid double indices. %In addition, it should be obvious that group-membership varies between subjects. 
These effects go one step further as they are not only heterogeneous and time-varying but the time-variation is potentially heterogeneous, too.
%As outlined before, the grouping cannot only correspond to treatments but also to other characteristics like gender or centers in multicenter studies. 

We model these effects by \emph{functional random coefficients}, which have been recently proposed by \citet{hagemann2024}. They are constructed by using \emph{functional random effects} \citep[FRE;][]{kneib2019} as varying coefficients. %Other authors, especially \citet{wood2017}, use the terms \emph{factor smooth interactions} or \emph{random wiggly curves} for such effects. 
For better readability, we will leave out the effect index $k$ for the remainder of this section. By using such FREs, we can model whole nonlinear time curves of continuous covariate effects group-specifically. 
FREs are essentially two-dimensional anisotropic tensor product interactions, that is $f_g(t):=f(g,t)$, of a random effect and a smooth time effect.  

In order to introduce anisotropic tensor product interactions formally, we first express the two main effects $f_1(g)$ and $f_2(t)$ in terms of basis function expansion as
$$
f_1(g)=\sum_{d_1=1}^{D_1}\gamma_{1d_1}B_{1d_1}(g), \quad f_2(t)=\sum_{d_2=1}^{D_2}\gamma_{2d_2}B_{2d_2}(t),
$$
where $B_{1d_1}(g)$ and $B_{2d_2}(t)$ are the basis functions, $\gamma_{1d_1}$ and $\gamma_{2d_2}$ are the basis coefficients and $D_1$ and $D_2$ are corresponding dimensions. For an introduction to univariate basis function expansions the reader is refereed to \citet{Fahrmeir2022} or \citet{wood2017}.
Their tensor product interaction is then given by 
$$f(g,t) = \sum_{d_1=1}^{D_1} \sum_{d_2=1}^{D_2} \gamma_{d_1d_2}B_{d_1d_2}(g,t),$$
where the \emph{tensor product basis functions} 
$$B_{d_1d_2}(g,t) = B_{1d_1}(g) B_{2d_2}(t)$$
result from pairwise interactions of the main effect basis functions.

While such tensor product interactions are mainly used to construct interaction surfaces of continuous variables, we can also use them to interact smooth effects with random effects by choosing the basis functions correspondingly. 
Hence, we consider i.i.d. random effects, i.e. log-normal frailty, for the first main effect. As outlined by \citet{kneib2019} and \citet{hagemann2024}, the corresponding basis function representation is given by
\begin{equation}
 f(g) = \sum_{d_1=1}^{D_1}\gamma_{1d_1} B_{1d_1}(g) =  \sum_{d_1=1}^{G}\gamma_{1d_1} \mathds{1}(g=d_1) = \gamma_{1g}
\end{equation}
with the $D_1=G$ basis functions being indicator functions for the group membership. 
The time-varying effect, i.e. the second main effect, can be modeled using P-splines \citep{eilers1996}. 
That is, using B-spline basis functions in combination with a discrete, usually a first- or second-order, penalty. Accordingly, the univariate penalty matrix of the random effect is given as a unit matrix $\boldsymbol{I}_G$ of dimension G and that of the P-spline as $\boldsymbol{D}_{D_2}^\top \boldsymbol{D}_{D_2}$ where $\boldsymbol{D}_{D_2}$ is a first or second order difference matrix of dimension $D_2$.
Alternatively, other forms of penalized splines can be used as well, e.g. thin plate splines \citep{wood2003}.

% \citet{kneib2019} propose a straight forward way to construct the tensor product penalties from Kronecker products of the two univariate penalty matrices. For a FRE that is 
% \begin{equation*}
%    \omega \boldsymbol{I}_{G} \otimes \boldsymbol{I}_{D_2}+ (1-\omega )(\boldsymbol{D}_{D_2}^\top \boldsymbol{D}_{D_2})\otimes \boldsymbol{I}_{G},
% \end{equation*}
% where $\omega \in (0,1)$ is an anisotropy parameter that needs to be estimated. Alternatively, one can also deploy two smoothness parameters, one for each direction of the tensor product, rather than one smoothness and one anisotropy parameter. This approach is originally a Bayesian one but can be transformed into a frequentist one similar to the way it is done in section 8.2.1 of \citet{Fahrmeir2022}.

There are different ways of implementing penalties for tensor product interactions: 
it can be based on a straight forward combination of univariate penalty matrices (see \citet{kneib2019} for such an approach). Alternatively, \citet{wood2013} developed an approach which is not such a straight forward combination of univariate penalty matrices but is numerically advantageous, especially when conducting frequentist inference.
%An alternative approach is developed by \citet{wood2013} which is not such a straight forward combination of univariate penalty matrices but has some numerical advantages, especially when conducting frequentist inference. 
This construction is based on reparameterizing the 
univariate smooths into fixed and random effects using an eigendecomposition of the penalty matrix. This leads to splitting the smooths into components that are not penalized, e.g. constant or linear terms, and components that are subject to penalization. The model matrix for the tensor product smooth is then constructed by calculating row-wise Kronecker products of these components. This results in each component being subject to at most one penalty which makes estimation numerically stable. For the detailed step-wise construction procedure see section 3 of \citet{wood2013}.

\section{Simulations} \label{sec:simu}

In this simulation study, we investigate the performance of the proposed approach with regard to two critical aspects: achieving superior fit when heterogeneous time-variation is present and preventing overfitting in its absence. 

\subsection{Software implementation} \label{sec:simu:software}
% Das kommt aus Sec. 2.1 und ist dort aktuell auskommentiert.
The data transformation discussed in Section \ref{sec:model:PEM} can be conducted in \texttt{R} by using the function \texttt{as\_ped} from the package \texttt{pammtools} \citep{bender2018b}. 

% Das kommt aus Sec. 2.3 und ist dort aktuell auskommentiert.
%All three inference methods introduced in Section \ref{sec:model:inference} are implemented in in the \texttt{R} package \texttt{mgcv}: Inference based on a smoothness selection criteria can be deployed by using the function \texttt{gam} with \texttt{method="GCV"}, a GLMM-based approach is implemented in the function \texttt{gamm} and the 
Frequentist estimation of the Poisson model is implemented in in the \texttt{R} package \texttt{mgcv}: the method of \citet{wood2011} can be used via the function \texttt{gam} with \texttt{method="REML"}.

%Due to the reasons outlined in the last paragraph of Section \ref{sec:model:inference}, we prefer to use the latter implementation, which is also suggested by \citet{bender2018a}. 

% Das kommt aus Sec. 3 und ist dort aktuell auskommentiert.
The tensor product constructor proposed by \citet{wood2013}, which is discussed in Section \ref{sec:FRE}, is implemented in the \texttt{R} package \texttt{mgcv} as function \texttt{t2}. In addition, based on \texttt{t2}, a numerically optimized implementation of the FRE is given by \texttt{s(bs = "fs")}. The FRE can then be deployed as varying coefficient by linearly interacting it via the \texttt{by} argument leading to a functional random coefficient. 
Using this numerically optimized version, we can implement a functional random coefficient for a variable \texttt{x} as
\begin{center}
    \texttt{s(g, t, by = x, bs = "fs", xt  = list(bs = "ps"), m = c(3, 1))},
\end{center}
where 
\texttt{g} is the grouping variable encoded as factor, \texttt{t} is the time variable and a cubic P-spline with first order penalty is used.

\subsection{Data generating processes and models} \label{sec:simu:dgp}

In order to generate survival data, both parts of the simulation study use the hazard function
\begin{equation} \label{eq:simu}
    \lambda_i(t|x_{1i}, x_{2i}) =
    \exp \left( 3t + 3x_{i1} + f(x_{i2}, t, g)\right),
\end{equation}
which depends on three explanatory variables: $x_{1}$, $x_{2}$ and $g$, the grouping variable with 4 levels.
In the first part, referred to as scenario (I), heterogeneous time variation is deployed for the effect of $x_2$, i.e. 
$
f(x_{2i}, t, g)=f_g(t) \cdot x_{2i}
$. We then compare the fit of the model using the functional random coefficient for $x_2$, referred to as model (i), to the one of competing models, which are given by model (ii) including heterogeneity (as random effect) and time-variation but not their interaction, model (iii) including only heterogeneity, and model (iv) including only time-variation. 
Hence, the three competitors contain less flexible nested effects and, therefore, are suspected to have an inferior model fit. 
In order to investigate the prevention of overfitting, the second part of the simulation study uses these three cases as data generating processes (DGP) and again fits the four models, leading to simulation scenarios (II)-(IV). 
The resulting four simulation scenarios are summarized in Table \ref{tab:scenarios}.

\begin{table}[b]
  \centering
  \caption{Overview over the four simulation scenarios, their DGPs and the models corresponding to these DGPs.}
    \begin{tabular}{ccc}
    \hline
    Scenario & Effect of $x_2$ in the DGP & Model \\
    \hline
    (I)   & Heterogeneous time-variation & (i) \\
    (II)  & Heterogeneity \& time-variation but no interaction & (ii) \\
    (III) & Heterogeneity only & (iii) \\
    (IV)  & Time-variation only & (iv) \\
    \hline
    \end{tabular}%
  \label{tab:scenarios}%
\end{table}%

\begin{figure}
    \centering
    \includegraphics[width=\linewidth]{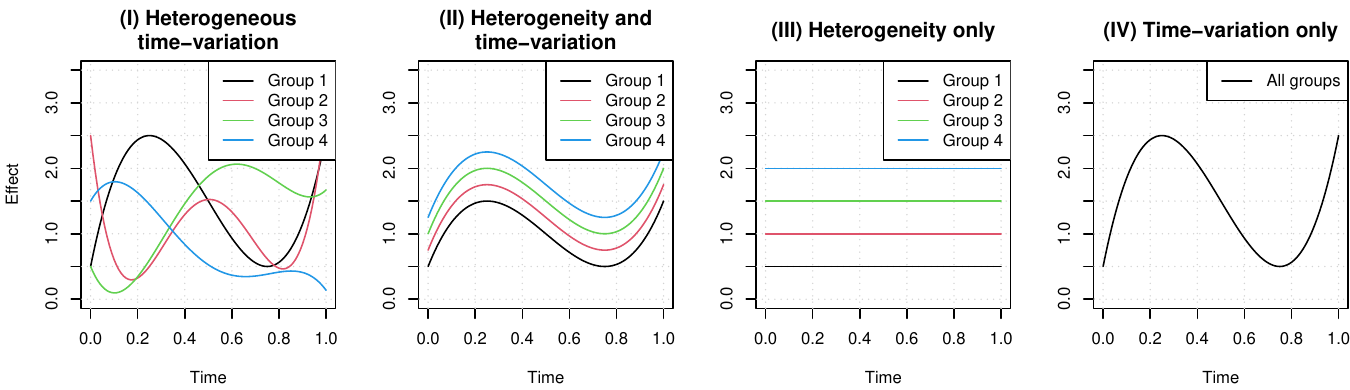}
    \caption{Effect of $x_2$ in the DGPs of the four scenarios. The first subplot (from the left) shows scenario (I), i.e. heterogeneous time-variation, the second one scenario (II), i.e. the combination of heterogeneity and time-variation (without interaction), the third one scenario (III), i.e. heterogeneity only, and the last one scenario (IV), i.e. time variation only.}
    \label{fig:dgp}
\end{figure}

Figure \ref{fig:dgp} shows the effect of $x_2$ that is deployed in the DGP of each of the four scenarios. It can be observed that in scenario (IV) all subjects follow the curve that applies for group 1 in scenario (I). The effect of scenario (II) results from adding up $\frac{1}{2}$ of the effect of scenario (III) and $\frac{1}{2}$ of the effect of scenario (IV). 

The simulation is conducted with
three different sample sizes, $n=200, 400$ and $800$ observations, equally distributed among the four groups. This leads to $50, 100$ and $200$ observations per group, respectively. The explanatory variables $x_1$ and $x_2$ are sampled from a uniform distribution, i.e.
$$
x_1, x_2 \stackrel{i.i.d.}{\sim} U[0,1]
$$
and the survival times are then sampled from \ref{eq:simu} using the algorithm of \citet{Bender2005}. Censoring is introduced with exponentially distributed censoring times, leading to an average censoring rate of $10.5 \%$. The in-sample model fit is evaluated based on the log-likelihood and the integrated Brier score \citep[IBS;][]{Graf1999}, also known as the cumulative in-sample prediction error. Hence, a smaller IBS is associated with a better model fit. The out-of-sample predictive accuracy is approximated based on an information criterion, namely the Akaike information criterion (AIC). The effective degrees of freedom, which is needed to compute the AIC of penalized models, is calculated according to \citet{Wood2016}.

\subsection{Results} \label{sec:simu:results}
\begin{figure}
    \centering
    \includegraphics[width=\linewidth]{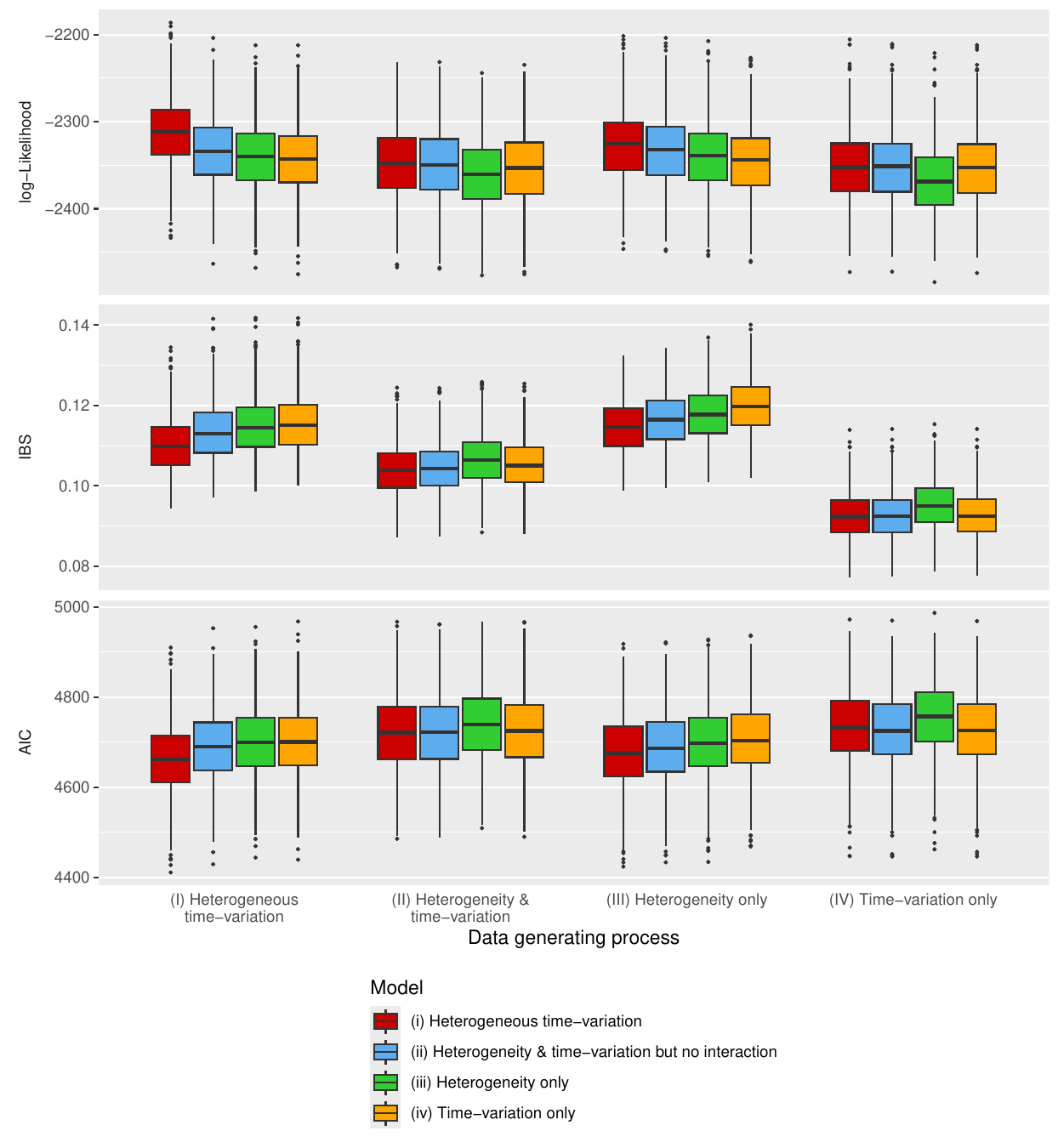}
    \caption{Results of the simulation study with $n=400$ in terms of the three fit measures. For each of the scenarios (I) - (IV) there is one block of consisting of four boxplots, one for each model (i) - (iv).}
    \label{fig:boxplots}
\end{figure}

The simulation is carried out with $1000$ repetitions 
and it can be observed that the results are almost the same across the different sample sizes.
Therefore, only the results for the medium sample size, which are shown in Figure \ref{fig:boxplots}, are discussed here. The outcomes for the small and large sample sizes are shown in Figures S1 and S2 of the supplementary materials.

It can be observed that, in the presence of heterogeneous time-variation, model (i) leads to a better model fit than the less flexible approaches (ii)-(iv). In addition, the difference between model (i) and the still quite flexible model (ii) is considerably larger than the difference between model (ii) and models (iii) and (iv). The fact that this also applies with regard to the AIC implies that this might not be caused by overfitting, but by modeling the underlying DGP more accurately. 

Regarding the fit of the models in scenarios (II) and (IV), 
%case that no heterogeneous time-variation is included tin the DGP, 
we can observe the desired behavior. % if the DGP includes heterogeneity and time-variation (but not their interaction) or time-variation only. 
Here, the fit of model (i) and the model resembling the DGP, i.e. model (ii) and (iv) respectively, is almost equal for all three fit measures. This strongly indicates that model (i) is penalized towards the true (less complex) model, as desired.
In contrast, in scenario (III) we can observe model (i) to fit slightly better than model (iii), which resembles the underlying DGP. However, when looking at the estimated models visually (visualizations for the first 100 simulation repetitions are uploaded together with the \texttt{R} code), it is, with few exceptions, clearly observable that there might most likely be no time-varying effect. In addition, it should be noted that the problem of slight overfitting also applies for model (ii).
%and, hence, might not be a inherent problem of the modeling approach.

In conclusion, the simulation study shows that the proposed functional random coefficient approach can flexibly capture heterogeneous time-variation within the covariable effects. In addition, if the time-variation is homogeneous, our proposed model does not lead to overfitting due being penalized towards the true model. Only if time-variation is fully absent, slight overfitting can be observed. However, in these cases, it is usually easy to visually recognize that there might not be any time variation in the data. Therefore, model selection should always involve a visual inspection and should not be based solely on fit measures. These results are mostly unaffected by the sample size.

\section{Brain tumor case study} \label{sec:case}
We apply the proposed approach to a brain tumor survival example based on data from \citet{Ceccarelli2016}. This dataset includes patients with a glioma divided into five different diagnoses: anaplastic astrocytoma, astrocytoma (other), anaplastic oligodendroglioma, oligodendroglioma (other), glioblastoma and mixed glioma. 
After removing 37 patients due to missing values, the total number of participants is $n=1094$, of which 593 are non-censored and 501 are censored. Survival times were recorded exact to the day and for 3 persons, who died already on their admission day, the survival time is set to half a day. We introduce an end-of-study, i.e. an administrative censoring, after 8 years because only 28 patients remain at this time.
In addition to the five diagnoses, the age and sex of the patients and \emph{fraction genome altered} (FGA) are recorded. 

%We aim to investigate the effect the \emph{fraction genome altered} (FGA) has onto the survival time. 
The FGA is commonly used in cancer research and represents the proportion of a tumor's genome that is affected by gains or losses of DNA segments, e.g. amplifications or deletions. 
Previous studies \citep[see, e.g.][]{Mehta2005} show that a higher FGA can be an indicator for aggressive tumor behavior and \citet{Dhital2023} even state that a high FGA is an independent predictor for a reduced overall survival.
%Besides, the FGA we include also the (sub-) diagnoses which are categorized into five groups: anaplastic Astrocytoma, Astrocytoma (other), anaplastic Oligodendroglioma, Oligodendroglioma (other), Glioblastoma and mixed Glioma. 
%In addition, we also include the patients sex and age as control variables. 

We suspect that the effect of the FGA might be time-varying and diagnosis-specific and that the time-variation might also be heterogeneous between diagnoses. Therefore, we model this effect by a functional random coefficient. This leads to the final model being a PAMM with a P-spline based baseline hazard, the diagnosis-specifically time-varying effect of FGA,  a linear effect for sex and age and a fixed effect for the diagnosis. For both, the log-baseline hazard and the functional random effect, we choose cubic P-splines with first order penalty and 9 inner knots, such that each of the intervals corresponds to one year. 

The estimated regression coefficients are shown in Table \ref{tab:coef}. 
A higher age significantly increases the hazard rate, which is an expected result as age usually increases the risk of death. In contrast, the effect of sex is not significant at a $5 \%$-level. Compared to the anaplastic astrocytoma, which is the reference category, all other diagnoses significantly influence the hazard function. While the risk of death is increased for the glioblastoma, it is reduced for the other three diagnoses. 

\begin{table}
  \centering
  \caption{Results for the fitted model: for the linear effects the estimated coefficients (Coef.), the standard error (SE) and the p-value, resulting from a z-test, is shown. For the smooth terms the p-value corresponds to the test of \citet{Wood2012}.}
  %estimated smoothing parameters (smooth. param.) and a p-value, corresponding to the test of \citet{}, is shown.}
    \begin{tabular}{lrrr}
    \hline
      \textbf{Linear effects:}   & Coef. & SE & p-value \\
    \hline
    Age   & 0.0401 & 0.0033 & $< 0.0001$ \\
    Sex: male & 0.1481 & 0.0849 & 0.0811 \\
    Diagnosis: astrocytoma (other) & -1.4830 & 0.4424 & 0.0008 \\
    Diagnosis: mixed glioma & -0.8081 & 0.2617 & 0.0020 \\
    Diagnosis: anaplastic oligodendroglioma & -0.8842 & 0.2868 & 0.0020 \\
    Diagnosis: oligodendroglioma (other) & -1.5816 & 0.3347 & $< 0.0001$ \\
    Diagnosis: glioblastoma & 0.9002 & 0.1670 & $< 0.0001$ \\
    \hline
    \textbf{Smooth terms:} & &      & p-value \\
    \hline
      log-baseline: $f(t)$    &   &       & $< 0.0001$ \\
      FGA: $f_{\text{Diagnosis}}(t)\cdot$ FGA    &    &  & 0.0002 \\
    \hline
    \end{tabular}%
  \label{tab:coef}%
\end{table}%

\begin{figure}
    \centering
    \includegraphics[width=\linewidth]{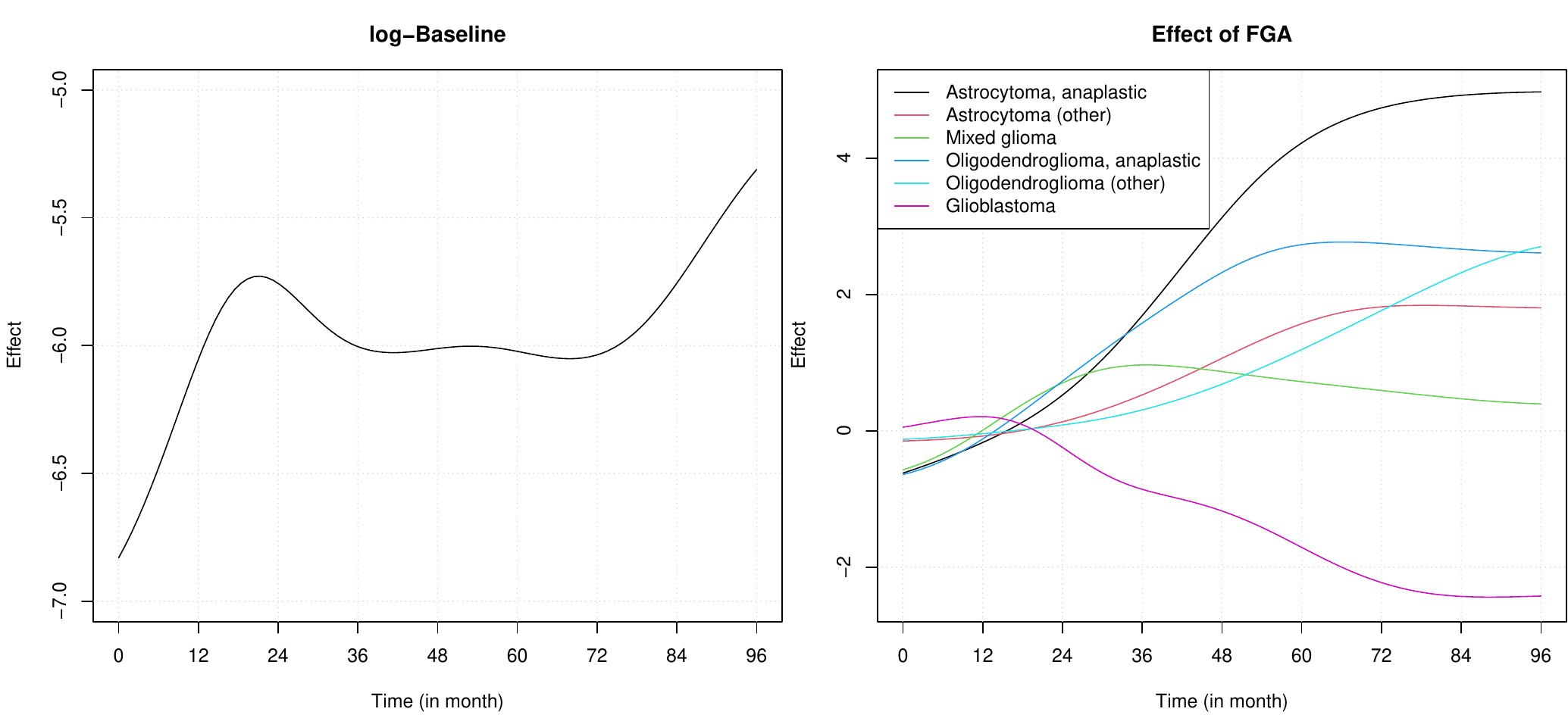}
    \caption{Estimated Smooth effects for the log-baseline function and the functional random coefficient of FGA.}
    \label{fig:case_study}
\end{figure}

The non-linear baseline hazard as well as the effect of the FGA is shown in Figure \ref{fig:case_study}. Both effects are significant with regard to the test of \citet{Wood2012}. 
The baseline hazard increases nearly linearly for the first year and a half, then decreases slightly for another year and a half, and then remains nearly constant for the next three years before increasing again. 
With regard to the effect of FGA, our main focus, we can indeed observe that the effect of FGA strongly varies over time as well as between the diagnoses and that the time variation is also quite different between the diagnoses. 
While there are no major differences in the first year and a half, the curves differ %significantly 
considerably thereafter. While the effect decreases over time for glioblastoma, it increases for the other four diagnoses. Hence, these effects might cancel out if the time-variation is not modeled diagnosis-specific. This outlines the practical relevance of our approach.

We compare the model fit to the competitors introduced in Section \ref{sec:simu} as well as a model deploying a simple linear effect of FGA, which is shown in Table \ref{tab:cs_fit}. 
It can be observed that modeling the time-varying effect of FGA diagnoses specifically leads to the best model fit. This also agrees with the visual impression from Figure \ref{fig:case_study}. 
Including only random effects or only time variation does not lead to a notable increase in model fit compared to the linear effect. In addition, even the model including heterogeneity as well as time-variation only slightly increases the model fit compared to the increase that is achieved by using the functional random coefficient. This coincides with the impression from Figure \ref{fig:case_study}, indicating that the time-variation might cancel out if it is not modeled diagnoses specifically. 
The fact that this also applies to the AIC indicates that this does not result from overfitting. 

% Table generated by Excel2LaTeX from sheet 'case_study_metrics'
\begin{table}
  \centering
  \caption{Fit measures for the proposed functional random coefficient and its competitors including a simple linear effect. The fit is measured in terms of the log-likelihood (logLik), the integrated Brier score (IBS) and the AIC.}
    \begin{tabular}{lccc}
    \hline      & logLik & IBS   & AIC \\
    \hline
    Heterogeneous time-variation   & -4099.36 & 0.1162 & 8238.87 \\
    Heterogeneity and time-variation   & -4108.27 & 0.1174 & 8250.03 \\
    Heterogeneity only & -4111.04 & 0.1185 & 8250.50 \\
    Time-variation only & -4111.00 & 0.1184 & 8251.44 \\
    Linear effect   & -4111.01 & 0.1184 & 8251.20 \\
     \hline
    \end{tabular}%
  \label{tab:cs_fit}%
\end{table}%

In conclusion, in this application, the use of a functional random coefficient allowed us to capture diagnosis-specific time variation in the effect of FGA on survival. 
In addition, the individual curves shown in Figure \ref{fig:case_study} can improve the understanding of FGA as a predictor of survival. 
%By modeling this effect explicitly, it is excluded from the residuals which also increases the model fit.
\section{Conclusion} \label{sec:conclusion}
In this paper, we introduced heterogeneously time-varying covariable effects to hazard regression models. This provides an appropriate model for cases in which the effect of a covariable is not only time-varying and subgroup-specific but its time-variation is subgroup specific, too. 

The proposed method makes use of the existing framework of PAMMs which enables us to deploy an efficient Poisson model-based inference.
%Our approach uses functional random coefficients and, hence, deploys an anisotropic tensor product interaction of a continuous time effect and frailty. 
Our approach allows for non-linear time-effects due to being based on penalized splines and uses an efficient random effects basis to model the heterogeneity. 
In addition, the penalization mostly prevents our method from overfitting in absence of heterogeneous time-variation. The corresponding simulation study only shows slight overfitting if time-effects are fully absent. However, it is easy to visually assess such cases. 
On the other hand, in presence of heterogeneous time-variation, the simulation study outlines the superior fit of your approach. 

We apply this model to a brain tumor case study. Here, the effect of the FGA varies over time and this time-variation is highly diagnosis-specific. Therefore, modeling FGA with a diagnosis-specific time-varying effect not only greatly improves the model fit, but also prevents the effects from canceling each other out. 
This provides additional interpretability and may lead to a better understanding of FGA as a risk predictor. Thus, this case study outlines the practical relevance of the proposed method.

Future possible research includes introducing this type of effect to Bayesian survival models. This is mainly a matter of computational efficiency since in a Bayesian setting, for both -- the piecewise exponential approach and the direct approach involving an integral over the hazard rate -- estimating a FRE might lead to very high computational costs. 
In addition, other applications, such as use in multicenter studies, should be explored. 

\section*{Supplementary Material}
Supplementary material is available online.

\section*{Software and data availability}
Software in the form of R code is available at \url{https://github.com/Niklas191/heterogeneous_time-variation}.
The case study data set is publicly available at the 
cBioPortal database with IDs \hyperlink{https://www.cbioportal.org/study/clinicalData?id=gbm_tcga_gdc}{gbm\_tcga\_gdc} and 
\hyperlink{https://www.cbioportal.org/study/clinicalData?id=difg_tcga_gdc}{difg\_tcga\_gdc}.
%\section*{Acknowledgements}
% \section*{Author contributions}

\section*{Funding}
This work has been supported by the Research Training Group "Biostatistical Methods for High-Dimensional Data in Toxicology`` (RTG 2624, P7) funded by the Deutsche Forschungsgemeinschaft (DFG, German Research Foundation, Project Number 427806116).

\section*{Competing interests}
The authors declare no competing interests.

%\bibliography{bib}

\begin{thebibliography}{}

\bibitem[Andersen and Gill, 1982]{andersen1982}
Andersen, P.~K. and Gill, R.~D. (1982).
\newblock Cox's regression model for counting processes: a large sample study.
\newblock {\em The annals of statistics}, pages 1100--1120.

\bibitem[Bender et~al., 2018]{bender2018a}
Bender, A., Groll, A., and Scheipl, F. (2018).
\newblock A generalized additive model approach to time-to-event analysis.
\newblock {\em Statistical Modelling}, 18(3-4):299--321.

\bibitem[Bender and Scheipl, 2018]{bender2018b}
Bender, A. and Scheipl, F. (2018).
\newblock pammtools: Piece-wise exponential additive mixed modeling tools.
\newblock {\em arXiv:1806.01042 [stat]}.

\bibitem[Bender et~al., 2005]{Bender2005}
Bender, R., Augustin, T., and Blettner, M. (2005).
\newblock Generating survival times to simulate cox proportional hazards
  models.
\newblock {\em Statistics in Medicine}, 24(11):1713--1723.

\bibitem[Ceccarelli et~al., 2016]{Ceccarelli2016}
Ceccarelli, M., Barthel, F.~P., Malta, T.~M., et~al. (2016).
\newblock Molecular profiling reveals biologically discrete subsets and
  pathways of progression in diffuse glioma.
\newblock {\em Cell}, 164(3):550--563.

\bibitem[Cox, 1972]{cox1972}
Cox, D.~R. (1972).
\newblock Regression models and life-tables.
\newblock {\em Journal of the Royal Statistical Society: Series B
  (Methodological)}, 34(2):187--202.

\bibitem[Dhital and Rodriguez-Bravo, 2023]{Dhital2023}
Dhital, B. and Rodriguez-Bravo, V. (2023).
\newblock Mechanisms of chromosomal instability (cin) tolerance in aggressive
  tumors: surviving the genomic chaos.
\newblock {\em Chromosome Research}, 31(2):15.

\bibitem[Eilers and Marx, 1996]{eilers1996}
Eilers, P. H.~C. and Marx, B.~D. (1996).
\newblock {Flexible smoothing with {B}-splines and penalties}.
\newblock {\em Statistical Science}, 11(2):89--121.

\bibitem[Fahrmeir et~al., 2022]{Fahrmeir2022}
Fahrmeir, L., Kneib, T., Lang, S., and Marx, B.~D. (2022).
\newblock {\em Regression: Models, Methods and Applications}.
\newblock Springer, Berlin, Heidelberg.

\bibitem[Friedman, 1982]{friedman1982}
Friedman, M. (1982).
\newblock Piecewise exponential models for survival data with covariates.
\newblock {\em The Annals of Statistics}, 10(1):101--113.

\bibitem[Graf et~al., 1999]{Graf1999}
Graf, E., Schmoor, C., Sauerbrei, W., and Schumacher, M. (1999).
\newblock Assessment and comparison of prognostic classification schemes for
  survival data.
\newblock {\em Statistics in Medicine}, 18(17-18):2529--2545.

\bibitem[Gray, 1992]{Gray1992}
Gray, R.~J. (1992).
\newblock Flexible methods for analyzing survival data using splines, with
  applications to breast cancer prognosis.
\newblock {\em Journal of the American Statistical Association},
  87(420):942--951.

\bibitem[Hagemann et~al., 2024]{hagemann2024}
Hagemann, N., Guhl, D., Kneib, T., M\"ollenhoff, K., and Steiner, W. (2024).
\newblock Dynamic heterogeneity in discrete choice experiments.
\newblock {\em Preprint available at SSRN: 4957076}.

\bibitem[Hennerfeind et~al., 2006]{hennerfeind2006}
Hennerfeind, A., Brezger, A., and Fahrmeir, L. (2006).
\newblock Geoadditive survival models.
\newblock {\em Journal of the American Statistical Association},
  101(475):1065--1075.

\bibitem[Hess, 1994]{Hess1994}
Hess, K.~R. (1994).
\newblock Assessing time-by-covariate interactions in proportional hazards
  regression models using cubic spline functions.
\newblock {\em Statistics in Medicine}, 13(10):1045--1062.

\bibitem[Hofner et~al., 2013]{hofner2013}
Hofner, B., Hothorn, T., and Kneib, T. (2013).
\newblock Variable selection and model choice in structured survival models.
\newblock {\em Computational Statistics}, 28:1079--1101.

\bibitem[Hofner et~al., 2011]{hofner2011}
Hofner, B., Kneib, T., Hartl, W., and K{\"u}chenhoff, H. (2011).
\newblock Building cox-type structured hazard regression models with
  time-varying effects.
\newblock {\em Statistical Modelling}, 11(1):3--24.

\bibitem[Jachno et~al., 2019]{Jachno2019}
Jachno, K., Heritier, S., and Wolfe, R. (2019).
\newblock Are non-constant rates and non-proportional treatment effects
  accounted for in the design and analysis of randomised controlled trials? a
  review of current practice.
\newblock {\em BMC Medical Research Methodology}, 19(1):103.

\bibitem[Kneib and Fahrmeir, 2007]{kneib2007}
Kneib, T. and Fahrmeir, L. (2007).
\newblock A mixed model approach for geoadditive hazard regression.
\newblock {\em Scandinavian Journal of Statistics}, 34(1):207--228.

\bibitem[Kneib et~al., 2019]{kneib2019}
Kneib, T., Klein, N., Lang, S., and Umlauf, N. (2019).
\newblock Modular regression - a lego system for building structured additive
  distributional regression models with tensor product interactions.
\newblock {\em TEST}, 28:1--39.

\bibitem[Li et~al., 2015]{Li2015}
Li, H., Han, D., Hou, Y., Chen, H., and Chen, Z. (2015).
\newblock Statistical inference methods for two crossing survival curves: a
  comparison of methods.
\newblock {\em PLoS One}, 10(1):e0116774.

\bibitem[Mehta et~al., 2005]{Mehta2005}
Mehta, K.~R., Nakao, K., Zuraek, M.~B., et~al. (2005).
\newblock Fractional genomic alteration detected by array-based comparative
  genomic hybridization independently predicts survival after hepatic resection
  for metastatic colorectal cancer.
\newblock {\em Clinical Cancer Research}, 11(5):1791--1797.

\bibitem[Murphy and Sen, 1991]{murphy1991}
Murphy, S.~A. and Sen, P.~K. (1991).
\newblock Time-dependent coefficients in a cox-type regression model.
\newblock {\em Stochastic Processes and their Applications}, 39(1):153--180.

\bibitem[Ripatti and Palmgren, 2000]{ripatti2000}
Ripatti, S. and Palmgren, J. (2000).
\newblock Estimation of multivariate frailty models using penalized partial
  likelihood.
\newblock {\em Biometrics}, 56(4):1016--1022.

\bibitem[Therneau et~al., 2003]{therneau2003}
Therneau, T.~M., Grambsch, P.~M., and Pankratz, V.~S. (2003).
\newblock Penalized survival models and frailty.
\newblock {\em Journal of Computational and Graphical Statistics},
  12(1):156--175.

\bibitem[Vaupel et~al., 1979]{vaupel1979}
Vaupel, J.~W., Manton, K.~G., and Stallard, E. (1979).
\newblock The impact of heterogeneity in individual frailty on the dynamics of
  mortality.
\newblock {\em Demography}, 16(3):439--454.

\bibitem[Wood, 2003]{wood2003}
Wood, S.~N. (2003).
\newblock Thin plate regression splines.
\newblock {\em Journal of the Royal Statistical Society Series B: Statistical
  Methodology}, 65(1):95--114.

\bibitem[Wood, 2011]{wood2011}
Wood, S.~N. (2011).
\newblock Fast stable restricted maximum likelihood and marginal likelihood
  estimation of semiparametric generalized linear models.
\newblock {\em Journal of the Royal Statistical Society: Series B (Statistical
  Methodology)}, 73(1):3--36.

\bibitem[Wood, 2012]{Wood2012}
Wood, S.~N. (2012).
\newblock On p-values for smooth components of an extended generalized additive
  model.
\newblock {\em Biometrika}, 100(1):221--228.

\bibitem[Wood, 2017]{wood2017}
Wood, S.~N. (2017).
\newblock {\em Generalized Additive Models: An Introduction with R, Second
  Edition}.
\newblock Chapman \& Hall/CRC Texts in Statistical Science. CRC Press.

\bibitem[Wood et~al., 2016]{Wood2016}
Wood, S.~N., Pya, N., and S\"afken, B. (2016).
\newblock Smoothing parameter and model selection for general smooth models.
\newblock {\em Journal of the American Statistical Association},
  111(516):1548--1563.

\bibitem[Wood et~al., 2013]{wood2013}
Wood, S.~N., Scheipl, F., and Faraway, J.~J. (2013).
\newblock Straightforward intermediate rank tensor product smoothing in mixed
  models.
\newblock {\em Statistics and Computing}, 23(3):341--360.

\bibitem[Zucker and Karr, 1990]{zucker1990}
Zucker, D.~M. and Karr, A.~F. (1990).
\newblock Nonparametric survival analysis with time-dependent covariate
  effects: a penalized partial likelihood approach.
\newblock {\em The Annals of Statistics}, 18(1):329--353.

\end{thebibliography}

\end{document}